\documentclass[aps,twocolumn,prb,superscript,floatfix,superscriptaddress,showpacs]{revtex4-1}
\usepackage{amssymb}

\usepackage{epsf}
\usepackage{epstopdf}
\usepackage{epsfig}
\usepackage{graphicx}
\usepackage{dcolumn}
\usepackage{braket}
\usepackage{bm}
\usepackage{amsfonts}
\usepackage{amsmath}
\usepackage{amssymb}
\usepackage{color,soul}
\usepackage{wasysym}
\usepackage{mathrsfs}
\usepackage{natbib}
\usepackage{multirow}
\usepackage[colorlinks=true,%
            linkcolor=blue,%
            urlcolor=blue,%
            citecolor=blue,%
            filecolor=blue,%
            bookmarksopen=true,%
            pdfauthor={J. P. Ramos Andrade},%
            pdftitle={SSinQDs},%
            pdfsubject={SSinQDs_Manuscript},%
            pdfpagemode=UseOutlines]{hyperref}

\newcommand{\ve}{\varepsilon}

\newcommand{\bea}{\begin{eqnarray}}
\newcommand{\eea}{\end{eqnarray}}

\newcommand{\la}{\langle}
\newcommand{\ra}{\rangle}

\begin{document}

\title{Spin-Seebeck effect and spin polarization in a multiple quantum dot molecule}
\author{J. P. Ramos-Andrade}
\email{juan.ramosa@usm.cl}	
\affiliation{Departamento de F\'isica, Universidad T\'ecnica Federico Santa Mar\'ia, Casilla 110 V, Valpara\'iso, Chile}
\author{F. J. Pe\~na}
\affiliation{Departamento de F\'isica, Universidad T\'ecnica Federico Santa Mar\'ia, Casilla 110 V, Valpara\'iso, Chile}
\author{A. Gonz\'alez}
\affiliation{Departamento de F\'isica, Universidad T\'ecnica Federico Santa Mar\'ia, Casilla 110 V, Valpara\'iso, Chile}
\author{O. \'Avalos-Ovando}
\affiliation{Department of Physics and Astronomy, and Nanoscale and Quantum Phenomena Institute, Ohio University, Athens, Ohio 45701-„1¤72979, USA}
\author{P. A. Orellana}
\affiliation{Departamento de F\'isica, Universidad T\'ecnica Federico Santa Mar\'ia, Casilla 110 V, Valpara\'iso, Chile}

\begin{abstract}
In this work, we study the conductance and the thermoelectric properties of a quantum dot embedded between two metallic leads with a side-coupled triple quantum dot molecule under a magnetic field. We focus on the spin polarization and thermoelectric quantities. Our results show the possibility of design an efficient spin filter device besides a noticeable enhancement of the Seebeck coefficient driven by the asymmetry in the quantum dots energy levels and a tunable pure spin Seebeck effect is obtained. This behavior also holds in the interacting case, where a pure spin Seebeck effect can be obtained for fixed values of the embedded quantum dot energy level. Our findings could lead to the implementation of a new  pure spin energy conversion and capable spin filter devices working with weak magnetic fields.

\end{abstract}
\date{\today}
\maketitle

\section{Introduction}

Thermoelectric phenomena characterize the ability of a system for converting waste of heat into electricity (or vice versa), when exposed to a temperature gradient across it (or voltage, when is the other way around). Nanoscale devices show enhanced thermoelectric properties concerning their macroscopic counterparts, allowing them to be very promising candidates for energy-efficient devices. While several studies have shown that a temperature gradient can produce electric currents, there are a few effects in where the same gradient can generate spin currents, better yet, in the total absence of electrical current, useful in efficient non-dissipative information processing. This new emerging area is known as spin caloritronics.\cite{Bauer2010,Bauer2012Spincaloritronics} A promising effect is the so-called spin Seebeck effect (SSE), which allows converting heat into a spin current (and vice versa via the inverse spin Hall effect\cite{Saitoh2006}), allowing to transport energy and information through materials such as ferromagnets, antiferromagnets, quantum dots (QDs), among others.

The SSE, first observed by Uchida \emph{et al.}\cite{uchida2008} in a ferromagnetic metal, has received a lot of attention over the last years.\cite{Uchida2010JAP,adachi2013} It has been observed in ferromagnetic,\cite{uchida2008} insulating,\cite{Uchida2010spin} semiconducting\cite{Jaworski2010} and conductive materials. More recently, the antiferromagnetic SSE has been predicted\cite{Rezende2016} and measured\cite{Wu2016} in MnF$_{2}$. Its enhancement has also been studied in heavy metals based hybrid structures,\cite{Guo2016} depending on thickness, temperature and interfacial effects. The SSE can be characterized by the spin-resolved Seebeck coefficient (or thermopower) $S$, when both spin components, $S_{\uparrow}$ and $S_{\downarrow}$, show equal magnitude and different signs, leading to the charge Seebeck coefficient vanishes ($S_{\text{c}}\propto S_{\uparrow}+S_{\downarrow}$) while the spin Seebeck coefficient is finite ($S_{\text{s}}\propto S_{\uparrow}-S_{\downarrow}$), producing the spin voltage at the time that the charge voltage is zero.\cite{Swirkowicz2009,Czerner2011,Trocha2012,Rameshti2015}

Although much SSE research has been done on ferromagnetic materials, less attention has been addressed to conductive ones, such as molecular junctions and QDs systems.\cite{Karwacki2016,Sierra2016} In these systems, attractive effects for achieving SSE are an enhanced thermoelectricity and spin polarization. Enhanced thermoelectric phenomena have been proved to exist if the transmission is strongly affected by quantum interference effects,\cite{Bergfield2009,Finch2009,Stadler2011,GomezSilva2012,Lambert2015} in particular the Fano effect.\cite{Miroshnichenko2010} On the other hand, the presence of a magnetic field or ferromagnetic leads is necessary for the thermoelectric response of the system becomes spin polarized. Pure spin Seebeck coefficients have been found for diverse setups, in where closed Aharonov-Bohm ring is one of the most studied, such as for Rashba QD molecules rings, \cite{Liu2011,gomez2014} for a degenerated molecular QD in a ring,\cite{Jiang2015} and for a mesoscopic ring in the presence of both Rashba and Dresselhaus spin–-orbit interactions.\cite{Liu2016} Coulomb correlation in a single QD in the presence of metal and magnetic insulator leads, has also been proved to tune and enhance the SSE, leading to an efficient thermovoltaic transistor.\cite{Lei2016} A T-shaped strongly coupled double QD system has been shown recently to exhibit SSE in the regime of the second stage of the Kondo effect.\cite{Krzysztof2016}

A similar effect, the so-called spin-current Seebeck effect, which is the charge voltage generation by a spin current, has been studied in graphene\cite{VeraMarun2012} and a single QD with electron-electron Coulomb interaction.\cite{Yang2014,Ramos2015} While most of the studies in QDs have focused in closed geometries such Aharonov-Bohm interferometers,\cite{Liu2011,gomez2014,Jiang2015} or single QDs in the presence of ferromagnetic leads,\cite{Krzysztof2016,Hwang2016} only a few have focused on the generation of spin currents of non-closed systems in the presence of metallic leads, systems such single QDs\cite{Dong2015} or lateral groups of multiple-QDs. A setup for the latter has been recently growth experimentally and an excitonic electrons attraction mediated by Coulomb repulsion was reported and schematized with a lateral group of multiple-QDs model.\cite{Hamo2016}

As multiple-QD systems have received increasing attention, since they have been proved to be useful setups to enhance thermoelectric quantities, \cite{Urban2015,Lambert2015,Li2016} this prompts the question of whether they are capable of achieving the SSE when connected to normal metallic leads, in the presence of a magnetic field. Previously, we studied a closed setup of a Rashba quantum dot molecule embedded in an Aharonov-Bohm interferometer, and we showed that SSE could be achieved, in the presence of normal metallic leads, by only suitable tuning the different phases.\cite{gomez2014} Later, we studied an open setup of a multi-QD system attached laterally to the conduction channel, or namely a side-QD molecule, for which we showed it could exhibit bound states in the continuum (BIC) and Fano effect in the absence of electron-electron interaction.\cite{ramos2014} In the present work, we study the creation of a pure SSE in the side-QD-molecule system connected to metallic leads, and the Fano effect role on the thermoelectric properties. We use a multi-impurities Anderson Hamiltonian to model the system, and the Green's functions formalism to calculate the transmission probability via the equation of motion (EOM) procedure and the thermoelectric quantities of the system. We also consider electron-electron interaction within Hubbard III approximation. We show that a tunable SSE can be achieved for different system temperatures. Moreover, we demonstrate that the SSE obtained in our system can be tuned via a  magnetic field and/or energy asymmetry of the external QDs for the non-interacting case. For the interacting case, in the Coulomb blockade regime, we find that a pure SSE can be achieved by fixing the energy level of the QD embedded between the two metallic leads and we show that the system can act as an efficient spin filter device.

The paper is arranged as follows: Section\ \ref{secmodel} shows the theoretical model, the electron-electron interaction approach, and the thermoelectric phenomena we are using. Section\ \ref{secresults} shows our results and the discussion, and\ \ref{secconclu} the concluding remarks of this work.

\section{Model}\label{secmodel}

\begin{figure}[tbph]
\centering
\includegraphics[width=6cm]{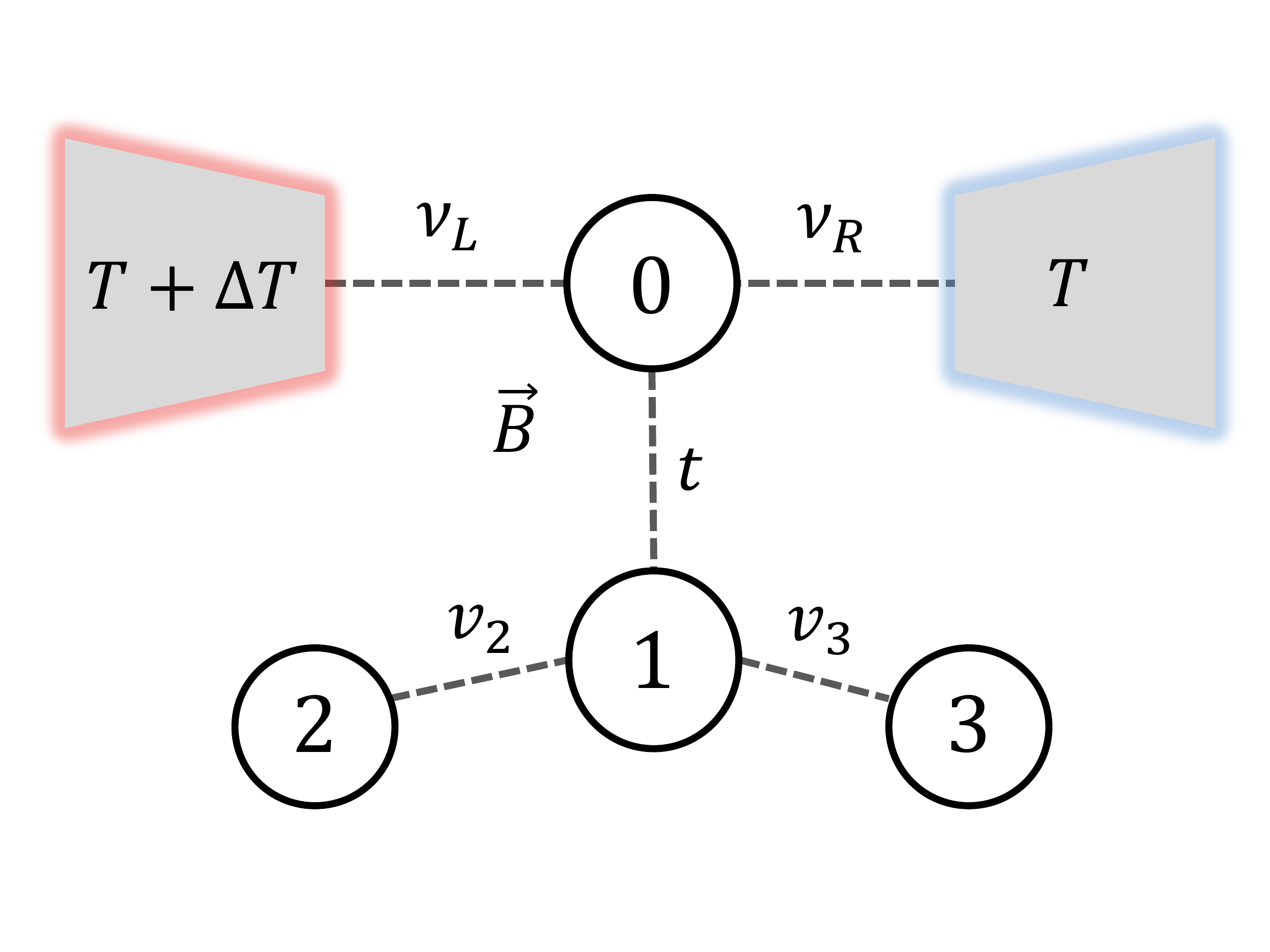}
\caption{Schematic view of our model: An embedded quantum dot between two metallic leads with a side-coupled triple-QD molecule. A magnetic field $\textbf{B}$ is set perpendicular to the plane. Both electrodes are given different temperatures, as shown.}
\label{Model}
\end{figure}

The system consists of single-level array of quantum dots coupled between them. One of them (QD0) is embedded between two leads at some temperature difference $\Delta T$. The others QDs (labelled as QD1, QD2 and QD3) form a side-molecule (QDM), which is side coupled to QD0, as shown in Fig.\ \ref{Model}. We model the system with an Anderson tunneling Hamiltonian within second quantization framework as follows
\begin{equation}
H=H_{\text{leads}}+H_{\text{dots}}+H_{\text{dot-leads}}\,, \label{H}
\end{equation}
where each contribution on the right side is given by
\begin{eqnarray}
H_{\text{leads}}&=&\sum_{k_{\alpha},\sigma}\ve_{k_{\alpha},\sigma}c_{k_{\alpha},\sigma}^{\dag}c_{k_{\alpha},\sigma}\,, \label{Hleads}\\
H_{\text{dots}}&=&\sum_{j=0,\sigma}^{3}\ve_{j,\sigma}d_{j,\sigma}^{\dag}d_{j,\sigma}+\sum_{\sigma}(t d_{0,\sigma}^{\dag}d_{1,\sigma}+\text{h.\,c.}) \nonumber \\
&+&Un_{0,\uparrow}n_{0,\downarrow}+\sum_{j=2,\sigma}^{3}(v_{j}d_{1,\sigma}^{\dag}d_{j,\sigma}+\text{h.\,c.})\,,\label{Hdots} \\
H_{\text{dot-leads}}&=&\sum_{k_{\alpha},\sigma}(\nu_{\alpha}c_{k_{\alpha},\sigma}^{\dag}d_{0,\sigma}+\text{h.\,c.}) \label{Hdotleads}\,,
\end{eqnarray}
where $d_{j,\sigma}^{\dag}$ ($d_{j,\sigma}$) is the electron creation (annihilation) operator in the $j$th (QD$j$) quantum dot with spin $\sigma=\,\uparrow\,\text{or}\,\downarrow$, $c_{k_{\alpha},\sigma}^{\dag}$ ($c_{k_{\alpha},\sigma}$) is the electron creation (annihilation) operator in the lead $\alpha=L,R$ with momentum $k$ and spin $\sigma$. $\nu_{\alpha}$ is the tunneling coupling between the embedded QD (QD0) and the lead $\alpha$. $U$ is the Coulomb's interaction in QD0, $t$ the tunneling coupling between QD0 and QD1, while $v_{2(3)}$ it is between the QD1 and the extremal QD2(3). In order to break the spin degeneracy, we include a magnetic field of magnitude $B$ to split each dot energy level by the Zeeman effect, as $\ve_{j,\sigma}=\ve_{j}+\sigma\ve_{z}$, where $\ve_{j}$ is the isolated dot energy level in the $j$th quantum dot, $\ve_{z}=g\mu_{\text{B}}B$ the Zeeman energy, being $g$ the land\'{e} factor and $\mu_{\text{B}}$ the Bohr magneton. Lastly, we adopted $\sigma=\,\uparrow(\downarrow)$ as $+1(-1)$.

We used the Green's function method as the starting point to obtain the transport quantities of interest by the EOM procedure. The retarded Green's function in the energy domain for the embedded quantum dot is given by (see Appendix)
\begin{eqnarray}
\la\la d_{0,\sigma},d_{0,\sigma}^{\dag} \ra\ra_{\ve} &=& \frac{1-\la n_{\bar{\sigma}}\ra}{\ve-\ve_{0,\sigma}+i\Gamma-\Sigma_{\text{QDM},\sigma}(\ve)} \nonumber \\ \label{green}
&+&\frac{\la n_{\bar{\sigma}}\ra}{\ve-\ve_{0,\sigma}-U+i\Gamma-\Sigma_{\text{QDM},\sigma}(\ve)}\,,
\end{eqnarray}
where $\Gamma$ is the coupling strength between both leads and the QD0, in the wide band limit. The function
\begin{equation}
\Sigma_{\text{QDM},\sigma}(\ve)=\frac{t^{2}\left[(\ve-\ve_{1,\sigma})^{2}-\Delta^{2}\right]}{(\ve-\ve_{1,\sigma})\left[(\ve-\ve_{1,\sigma})^{2}-\Delta^{2}-2v^{2}\right]}\,, \label{Sigma}
\end{equation}
is the self-energy due to the lateral coupling of the molecule with QD0, where we have defined $v_{2}=v_{3}\equiv v$ and set $\ve_{2(3),\sigma}\equiv\ve_{1,\sigma}+(-)\Delta$, being $\Delta$ an asymmetric energy parameter for the external dots, and $\la n_{\bar{\sigma}}\ra$ is the occupation number with spin $\bar{\sigma}$, opposite to $\sigma$, given by
\begin{equation}
\la n_{\sigma} \ra = -\frac{1}{\pi}\int f(\ve,\mu)\,\text{Im}\left[\la\la d_{0,\sigma},d_{0,\sigma}^{\dag} \ra\ra_{\ve}\right]\,\text{d}\ve\,, \label{n}
\end{equation}
with $f(\ve,\mu)$ the Fermi distribution function. We calculate the transmission probability across the embedded quantum dot using
\begin{equation}
\mathcal{T}_{\sigma}(\ve)=-\Gamma\,\text{Im}\left[\la\la d_{0,\sigma}^{\dag}d_{0,\sigma}\ra\ra_{\ve}\right]\,, \label{tras}
\end{equation}
which final analytical expression can be found at the end of the Appendix section.
For the thermoelectric properties, we consider a system in linear response regime. In this regime, we can write the charge and heat current, $I_{\text{charge}}$ and $I_{\text{heat}}$ respectively, in terms of a potential difference $\Delta V$ and a temperature difference $\Delta T$ between the two leads, as\cite{mahanbook}
\begin{eqnarray}
I_{\text{charge}}&=&-e^{2}L_{0,\sigma}\Delta V+\frac{e}{T}L_{1,\sigma}\Delta T\,, \label{Icarga} \\
I_{\text{heat}}&=&e L_{1,\sigma}\Delta V-\frac{1}{T}L_{2,\sigma}\Delta T\,, \label{Iheat}
\end{eqnarray}
being $T$ the absolute temperature, $e$ the electron charge and the kinetic coefficients are given by
\begin{equation}
L_{n,\sigma}(\mu)=\frac{1}{h}\int\left(-\partial f(\ve,\mu)/\partial\ve\right)(\ve-\mu)^{n}\mathcal{T}_{\sigma}(\ve)\text{d}\ve\,,
\end{equation}
where $\mu$ is the chemical potential and $h$ the Planck's constant. The spin-dependent electrical conductance is obtained directly from above as $G_{\sigma}(\mu)=e^{2}L_{0,\sigma}(\mu)$. The Seebeck coefficient $S$ is the proportionality between the temperature difference $\Delta T$ and the potential difference $\Delta V$ caused when the charge current vanishes, so it is defined, per spin, as
\begin{equation}
S_{\sigma}(\mu)=-\frac{\Delta V}{\Delta T}=-\frac{1}{e\,T}\frac{L_{1,\sigma}}{L_{0,\sigma}}\,. \label{S}
\end{equation}

We define the charge-Seebeck coefficient $S_{\text{c}}(\mu)$ and spin-Seebeck coefficient $S_{\text{s}}(\mu)$ respectively as\cite{Swirkowicz2009,Czerner2011,Trocha2012,Rameshti2015}
\begin{equation}
S_{\text{c}(\text{s})}(\mu)\equiv S_{\uparrow}(\mu)+(-)S_{\downarrow}(\mu)\,. \label{Scargayespin}
\end{equation}

In order to explore the ability of the system to work as a spin filter, we consider the weighted polarization, defined as
\begin{equation}
P_{\sigma}=\frac{G_{\sigma}}{G_0}\left(\frac{G_{\uparrow}-G_{\downarrow}}{G_{\uparrow}+G_{\downarrow}}\right)\,,
\end{equation}
where $G_0=e^{2}/h$.

\section{Results}\label{secresults}

In what follows, we use $\Gamma=1$ meV as energy unit, and the coupling between QD1 with QD2(3) is fixed as $v=\Gamma/2$. For instance, according to this the Zeeman energy $\ve_{z}=0.1\,\Gamma$ corresponds to a magnetic field $B\sim 1$ T. The first is a suitable choice since for experimental targets of our system, one could consider QDs where $\Gamma=5$ meV\cite{Kobayashi2002} or molecular junctions where $\Gamma=2.1-5$ meV.\cite{Capozzi2016} Besides, we fixed the molecule QDs energy level to be resonant with the Fermi energy; this is $\ve_{\text{F}}\equiv0$. Lastly, we note that the proposed effects presented here could be achieved in laboratories by measurements of thermoelectric quantities\cite{Reddy2007,Widawsky2012,Nichele2015,Dutta2017} and by the suitable combination of tunable parameters such as small magnetic fields,\cite{Nichele2015,Chengyu2017} Coulomb interactions,\cite{Hamo2016} and temperatures.\cite{Chengyu2017,Dutta2017}

\subsection{Non-interacting case: $U=0$.}

First, we show the results for a vanishing Coulomb interaction in QD0, so the spin degeneracy breaking is due completely to the magnetic field via Zeeman effect. As a way of comparison, we start without magnetic field and then we will discuss the spin dependent case.

An important point of this work focuses on the effect of a side coupled QDM on $S$. The case without the magnetic field, spin-independent ($\ve_{z}=0$), can be explored for two scenarios: 
First, by changing the hopping parameter $t$, for the symmetric case, i.e., all the extremal QDs (2 and 3) have the same energy $(\Delta=0)$; and second, by fixing the hopping parameter $t$ and changing the asymmetric energy parameter of external QDs $\Delta$.

\begin{figure}[tbph]
\centering
\includegraphics[width=0.475\textwidth]{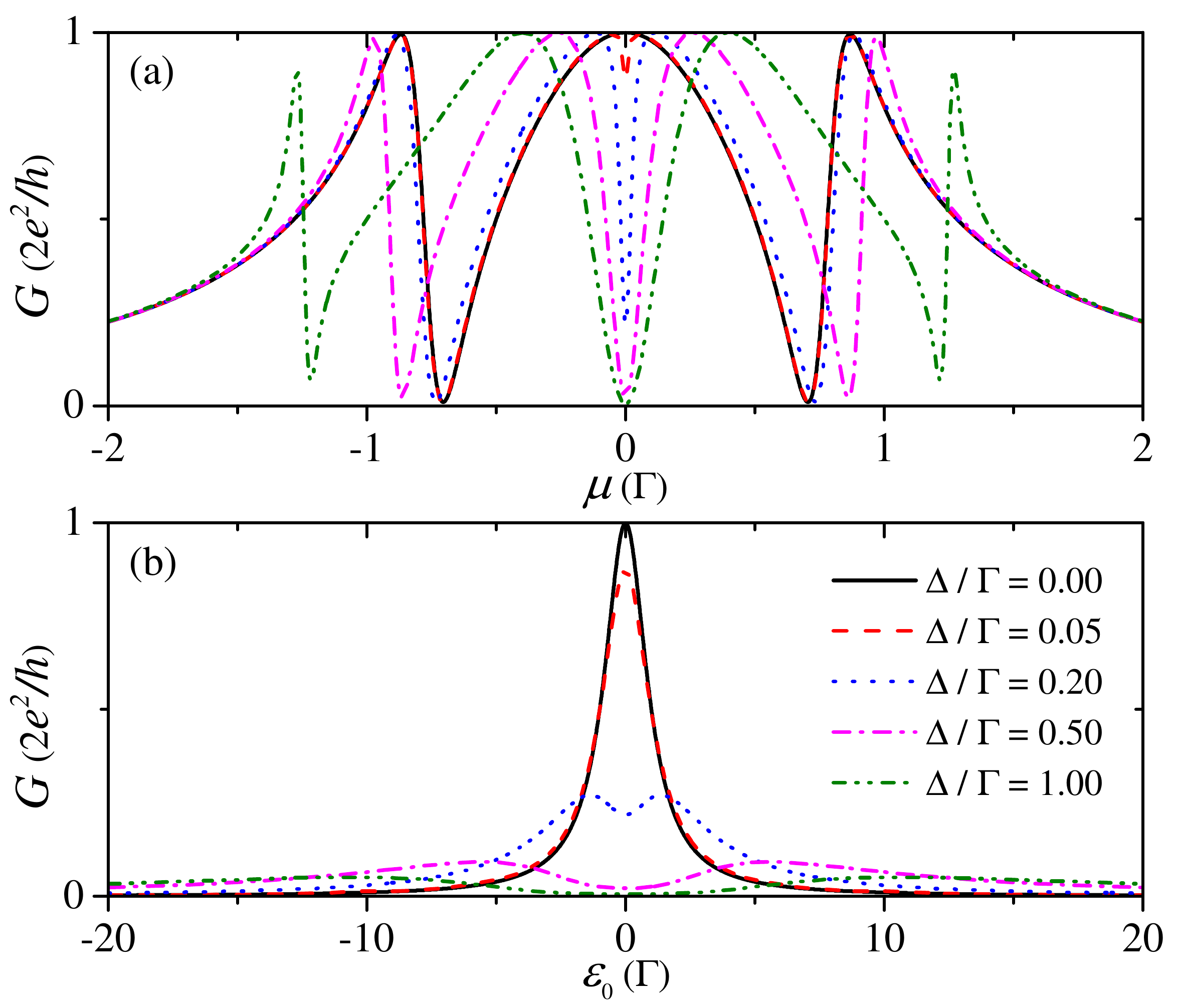}
\caption{Conductance $G$ for $k_{\text{B}}T/\Gamma=6.89$, for different values of $\Delta$, as a function of $\mu$ for fixed $\ve_{0}=0$ (a) and $\ve_{0}$ for fixed $\mu=0$ (b). Here $t=\nu=\Gamma/2$.}
\label{Fig2}
\end{figure}

\begin{figure}[tbph]
\centering
\includegraphics[width=0.475\textwidth]{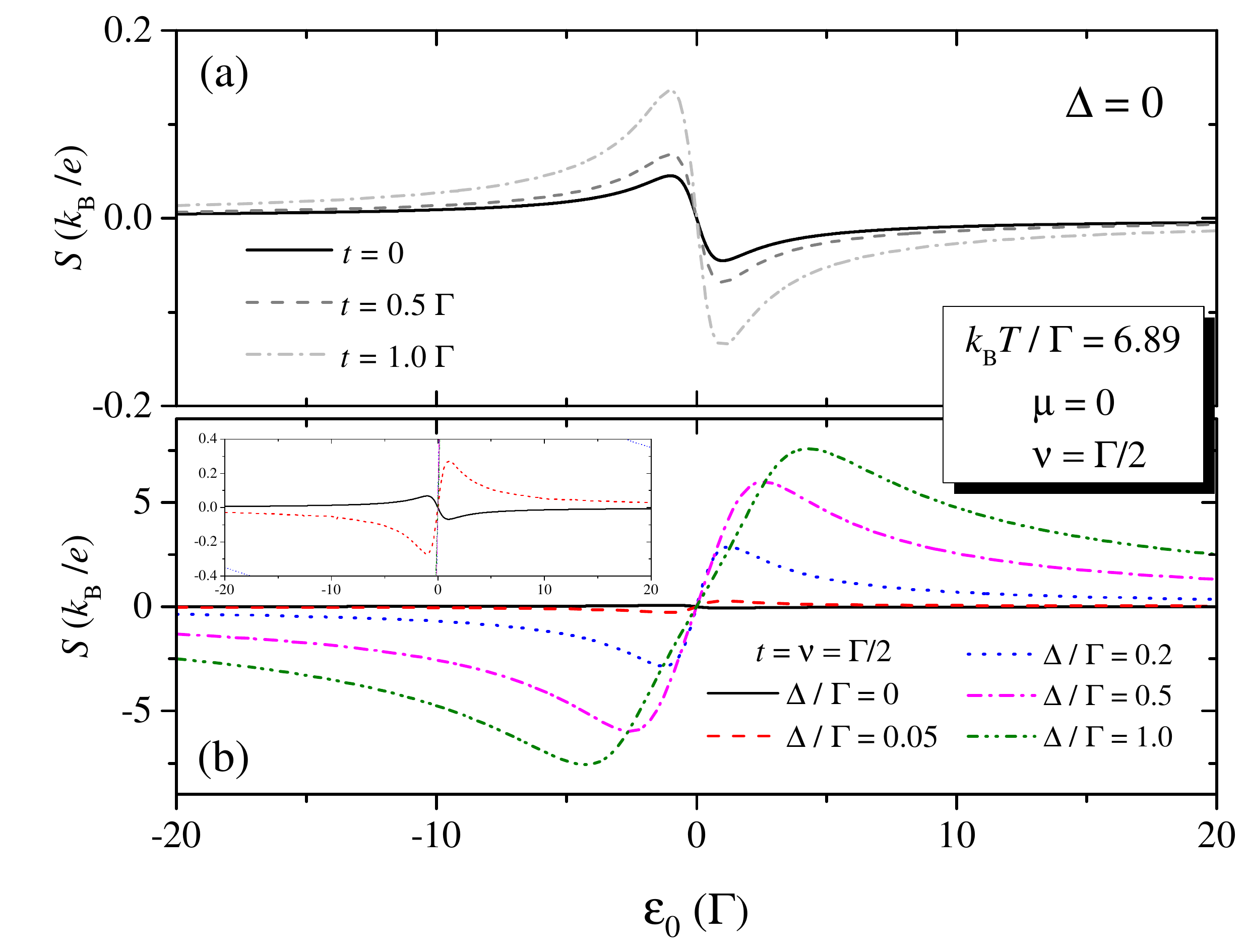}
\caption{Seebeck coefficient $S$ as a function of $\ve_{0}$ using $\Delta=0$: (a) for different $t$, and (b) fixed $t=\Gamma/2$ for different $\Delta$. Here the temperature is fixed to $k_{\text{B}}T/\Gamma=6.89$.}
\label{Fig3}
\end{figure}

We begin by discussing the conductance shape. Figure\ \ref{Fig2} shows the effect of the asymmetry parameter $\Delta$ on the conductance for the case without magnetic field $(\ve_{z}=0)$. First, from Fig.\ \ref{Fig2}(a), for $\Delta=0$ (black solid line) we observe two Fano antiresonances in the conductance as a function of $\mu$, which arise from the interference between continuum states of the embedded QD and localized states in the side-coupled molecule. Turning on $\Delta\neq0$ (non-solid lines) the conductance displays peaks due to the degeneracy-splitting of the resonances, hence contributing with additional Fano line shapes. This feature has a notorious consequence in Fig.\ \ref{Fig2}(b), which is to decrease the conductance amplitude as a function of the QD0 energy level, which is proportional to the Onsager coefficient $L_{0}$. So, for weak changes in the Onsager coefficient $L_{1}$, the net effect of the asymmetry parameter should be the enhancement of the Seebeck coefficient, because it is proportional to the ratio $L_{1}/L_{0}$. From Fig. \ref{Fig3}(a), we note the effect on $S$ for different values of the coupling tunneling  between QD0 and QD1, for vanishing $\Delta$. We note that for larger values of $t$ we obtain an increasing value of $S$, but to obtain a significant Seebeck coefficient value, or a so-called \emph{good Seebeck}\cite{Gabi2015,Chen2012,isern2016} $(S\geq 1$ in units of $k_{\text{B}}/e$), we need a value of $t\approx 6\,\Gamma$. On the other hand, by taking advantage of the asymmetry parameter dependence on the conductance shape, in the Fig.\ \ref{Fig3}(b) we plot $S$ for a fixed value of $t$ and different $\Delta$, observing a substantial enhancement of the Seebeck coefficient when $0<\Delta\leq\Gamma$. It is important to note that the more asymmetry in the external QDs there is, the more notorious the Seebeck we obtain. Besides, from the inset in Fig.\ \ref{Fig3}(b), in addition to the change in amplitude for $S$ by tuning $\Delta=0$ to $\Delta\neq 0$, we observe a sign inversion in $S$.

\begin{figure}[tbph]
\centering
\includegraphics[width=0.475\textwidth]{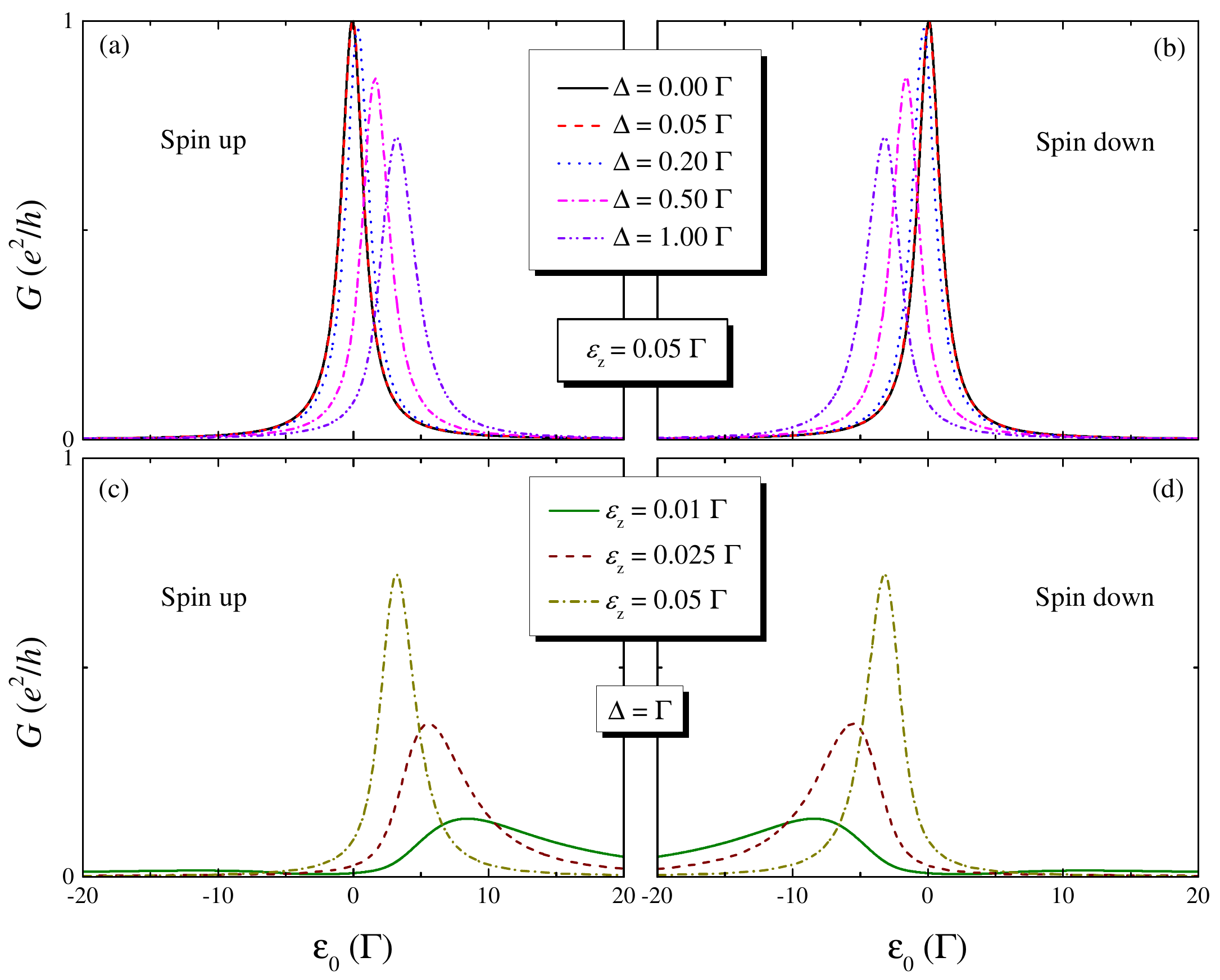}
\caption{Conductance $G$ for spin up (a) and down (b) for different values of $\Delta$ and fixed $\ve_{z}=0.05\,\Gamma$. Conductance $G$ for spin up (c) and down (d) for different values of $\ve_{z}$ and fixed $\Delta=\Gamma$. In all panels $k_{\text{B}}T/\Gamma=6.89$, $\mu=0$ and $t=\Gamma/2$.}
\label{Fig4}
\end{figure}

When a magnetic field is turned on, the spin-dependent conductance takes the form shown in the Fig.\ \ref{Fig4}. From it, we can understand the effect and the role of the self-energy $\Sigma_{\text{QDM},\sigma}(\ve)$. For the disconnected case ($t=0$), we expect a maximum in the conductance in a point near to $\ve_{0}+(-)\ve_{z}= \ve_{\text{F}}$, so for our case $(\ve_{\text{F}}\equiv 0)$, that point must be close to $\ve_{0}\approx -(+)\ve_{z}$. Then, the single peak in the conductance for the disconnected case for spin up [down] must be located on the left [right] in Fig.\ \ref{Fig4} panel (a) [panel (b)]. Hence, the effect of the side-coupled QDM ($t=\Gamma/2\neq 0$), is to provide an additional shift of the conductance peak $\Sigma_{\text{QDM},\uparrow(\downarrow)}$ to the right (left) for spin up (down). Note that the conductance amplitude decreases as $\Delta$ increases for fixed Zeeman energy,  as a consequence of the finite temperature in the system and the increment of $\Sigma_{\text{QDM},\sigma}$, which is in agreement with the case without magnetic field. On the other hand, panels (c) and (d) of Fig.\ \ref{Fig4} show another remarkable and counterintuitive effect, which is the increasing of $\Sigma_{\text{QDM},\sigma}$ as the magnetic field decreases with fixed $\Delta$. Bringing the consequence of the decreasing in conductance amplitude and also a larger spin up-down peak shifting with smaller magnetic fields. 

\begin{figure}[tbph]
\centering
\includegraphics[width=0.45\textwidth]{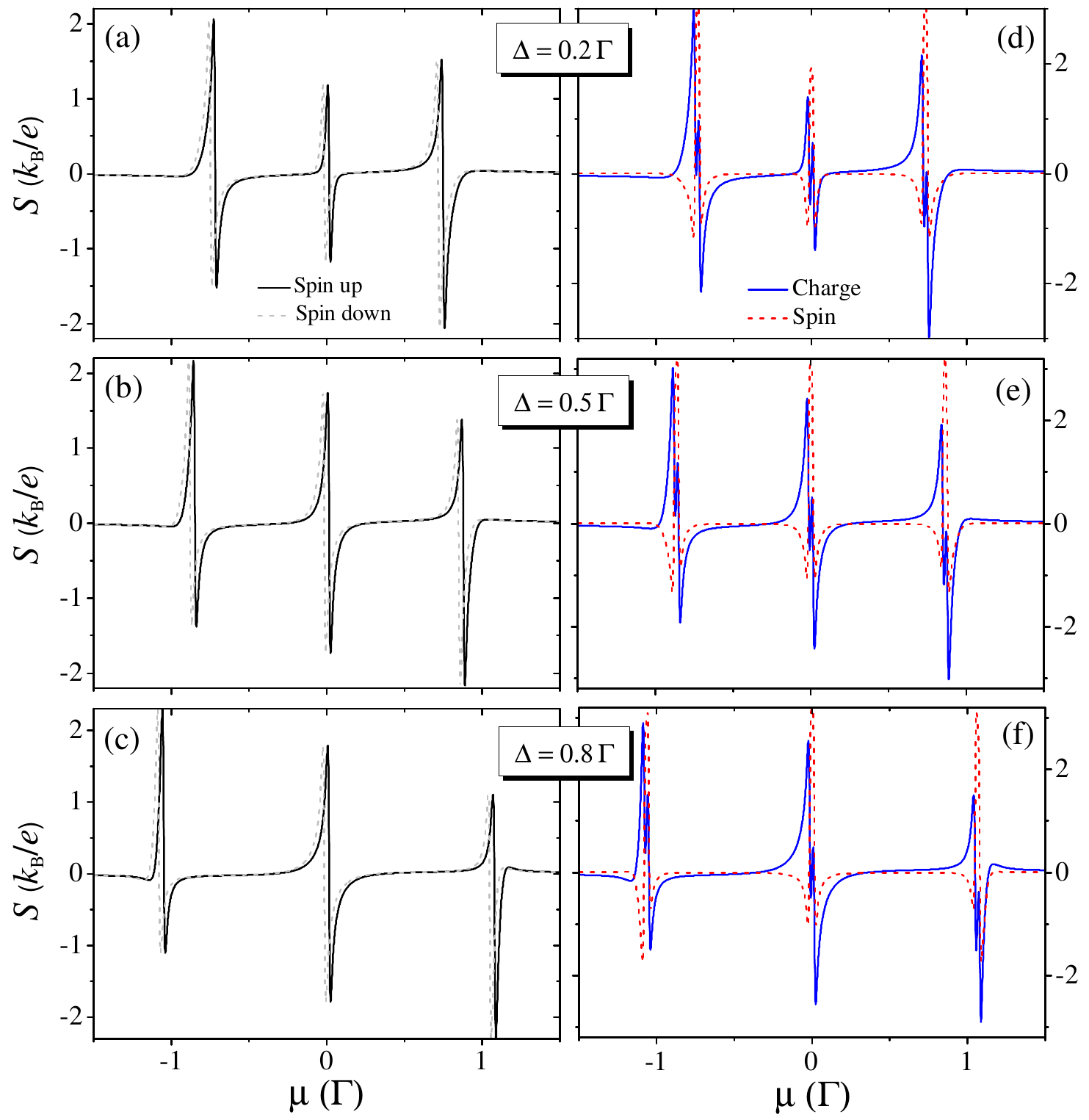}
\caption{Seebeck coefficient $S$ as a function of $\mu$ for fixed $\ve_{z}=0.015\,\Gamma$ and different $\Delta$: $\Delta=0.2\,\Gamma$ [(a) and (d)]; $\Delta=0.5\,\Gamma$ [(b) and (e)] and $\Delta=0.8\,\Gamma$ [(c) and (f)]; Left panels are for spin-resolved $S$ while right panels are for charge and spin $S$. Here $k_{\text{B}}T/\Gamma=5.17$, $t=\Gamma/2$ and $\ve_{0}=0$.}\label{Fig5}
\end{figure}

\begin{figure}[tbph]
\centering
\includegraphics[width=0.475\textwidth]{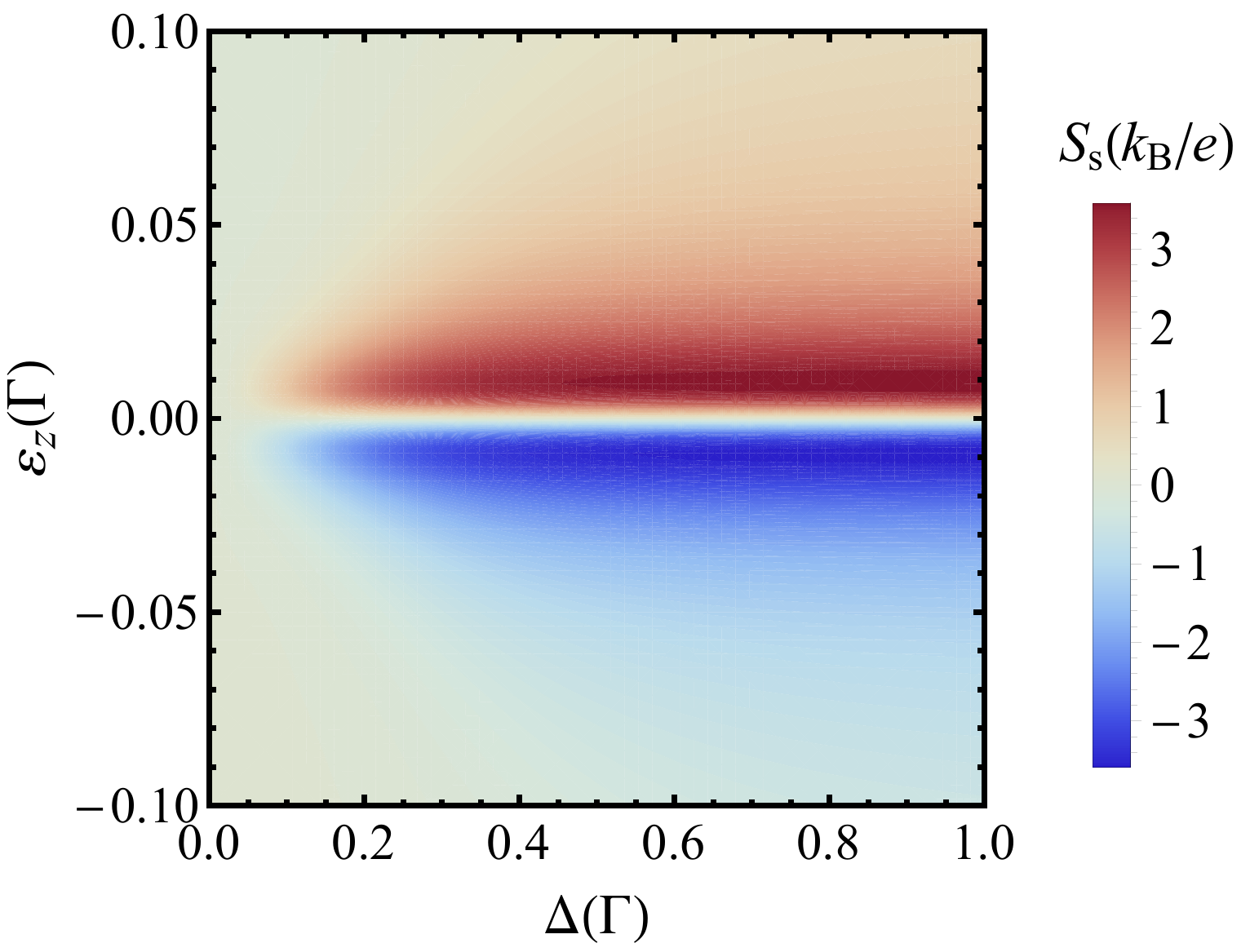}
\caption{Spin Seebeck coefficient $S_{\text{s}}$ as a function of $\ve_{z}$ and $\Delta$, for $\mu=0$, $k_{\text{B}}T/\Gamma=5.17$, $t=\Gamma/2$ and $\ve_{0}=0$.}
\label{Fig6}
\end{figure}

\begin{figure}[tbph]
\centering
\includegraphics[width=0.475\textwidth]{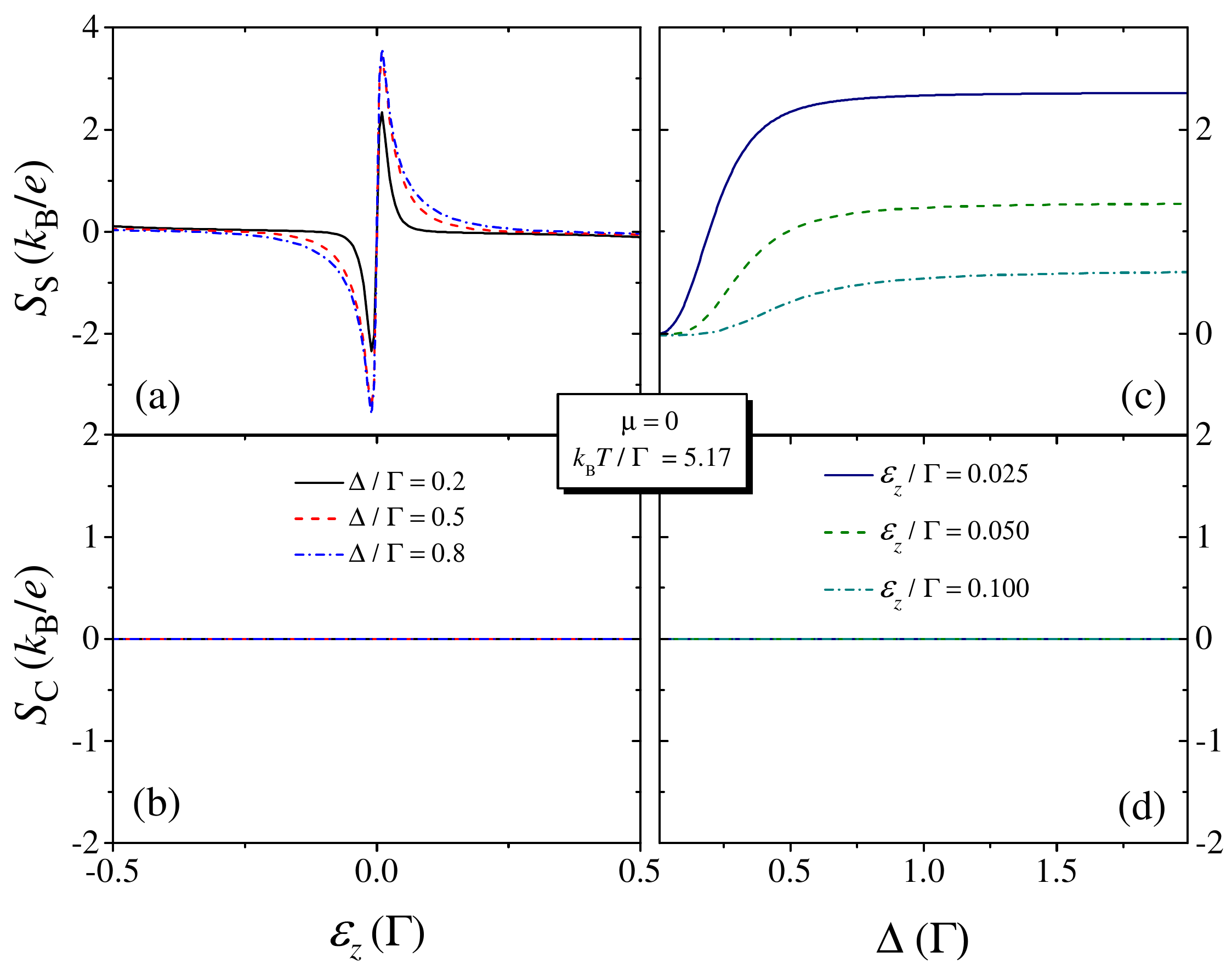}
\caption{Spin-Seebeck $S_{\text{s}}$ and charge-Seebeck $S_{\text{c}}$ coefficients as a function of $\ve_{z}$ [(a) and (b), respectively] and $\Delta$ [(c) and (d), respectively]. Here $k_{\text{B}}T/\Gamma=5.17$, $t=\Gamma/2$ and $\mu=\ve_0=0$.}
\label{Fig7}
\end{figure}

On the left panels of Fig.\ \ref{Fig5} we can observe the splitting of $S$ in $S_{\uparrow}$ and $S_{\downarrow}$ as a function of the chemical potential due to the Zeeman effect. Moreover, on the right panels of Fig.\ \ref{Fig5} an important feature in our model arises. For a fixed $\mu=0$ the charge-Seebeck coefficient always vanishes while the spin-Seebeck is non-zero, \emph{i. e.} $S_{\uparrow}(\mu=0)=-S_{\downarrow}(\mu=0)\neq0$. To clarify this, we show a contour plot of the spin-Seebeck in Fig.\ \ref{Fig6}, in the space of parameters $\ve_{z}$ and $\Delta$, using $\mu=0$ and $\ve_{0}=0$. For this range of parameters $S_{\text{c}}$ always vanishes (not shown), whereas $S_{\text{s}}$ is finite and non-zero in the most of the area in this domain. Therefore, we have a pure spin-Seebeck effect for fixed $\mu=0$. Besides, we can observe a special zone in Fig.\ \ref{Fig6} in which the spin-Seebeck maximizes, taking place for low magnetic fields and large $\Delta$. To explore this, we show the behavior of both $S_{\text{c}}$ and $S_{\text{s}}$ [Eq.\ (\ref{Scargayespin})] as a function of the Zeeman energy for three fixed $\Delta$ values in Fig.\ \ref{Fig7} panel (a), in which we note that the amplitude and sign of $S_{\text{s}}$ can be controlled via tuning the direction and/or amplitude of the magnetic field. In Fig.\ \ref{Fig7} panel (c) we plot $S_{\text{s}}$ as a function of $\Delta$ for three different Zeeman energies, where we obtain an asymptotic behavior of $S$ for large $\Delta$ values, \emph{i. e.} $\Delta\gtrsim\Gamma$. We emphasize that both results correspond to a pure spin-Seebeck effect since panels (b) and (d) show the vanishing of $S_{\text{c}}$ and that we can obtain a sizable spin Seebeck coefficient value with a weak Zeeman splitting, meaning a weak magnetic field.

\begin{figure}[tbph]
\centering
\includegraphics[width=0.475\textwidth]{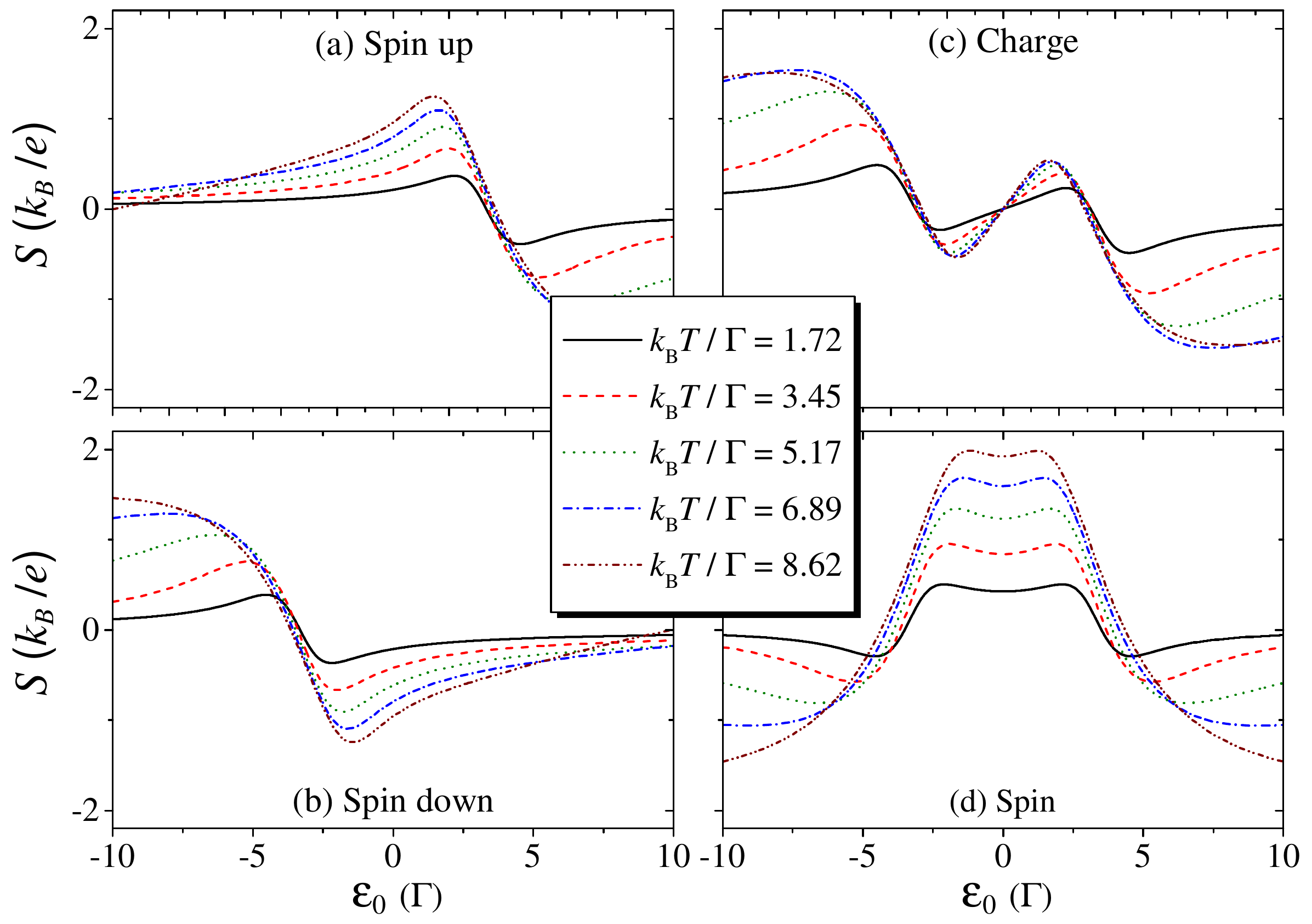}
\caption{Seebeck coefficient $S$ as a function of $\ve_{0}$ using $\ve_{z}=0.05\,\Gamma$ and $\Delta=\Gamma$. (a) and (b) for spin up and down respectively; (c) and (d) for charge and spin.}
\label{Fig8}
\end{figure}

In Fig.\ \ref{Fig8} we show the Seebeck coefficient as a function of the embedded dot energy level ($\ve_{0}$), which can be seen as a tunable gate potential. Panels (a) and (b) contain spin-resolved, while panels (c) and (d) show the charge and spin Seebeck coefficients respectively. In Fig.\ \ref{Fig8} panel (c) we observe a linear behavior of $S_{\text{c}}$ around $\ve_{0}=0$ with $S_{\text{c}}(\ve_{0}=0)=0$. This allows for a pure SSE to appear, since there exists a $S_{\text{s}}(\ve_0=0)\neq 0$, increasing its value with the temperature, as shown in panel (d).

\subsection{Interacting case: finite $U$}

In this subsection, we will analyze the system by including interaction in the QD0. We emphasize that we are interested in a Coulomb blockade regime in which charge fluctuations are the dominant events in transport. Hence, the Green's function (see Appendix) can be taken within the Hubbard III approximation, which corresponds to the most suitable description for the electronic correlations in the Coulomb blockade scenario, given by $U\gg\Gamma$. \cite{Hubbard401,wir2009,rejec2012} This kind of approach yields an excellent characterization of the Seebeck coefficient and the conductance of strongly interacting QDs for temperatures higher than the Kondo temperature\cite{Haldane1978} or when the localized level is weakly coupled to electrodes.\cite{isern2016} Throughout this section, we use the chemical potential fixed at $\mu=0$ and $U=10\,\Gamma$ for the Coulomb interaction, locating the system in the so-called Coulomb blockade regime. It is worth to mention that in this work, according to the parameters, Kondo temperature is approximately $k_{\text{B}}T_{\text{K}}\sim 0.05\,\Gamma$.

\begin{figure}[tbph]
\centering
\includegraphics[width=0.45\textwidth]{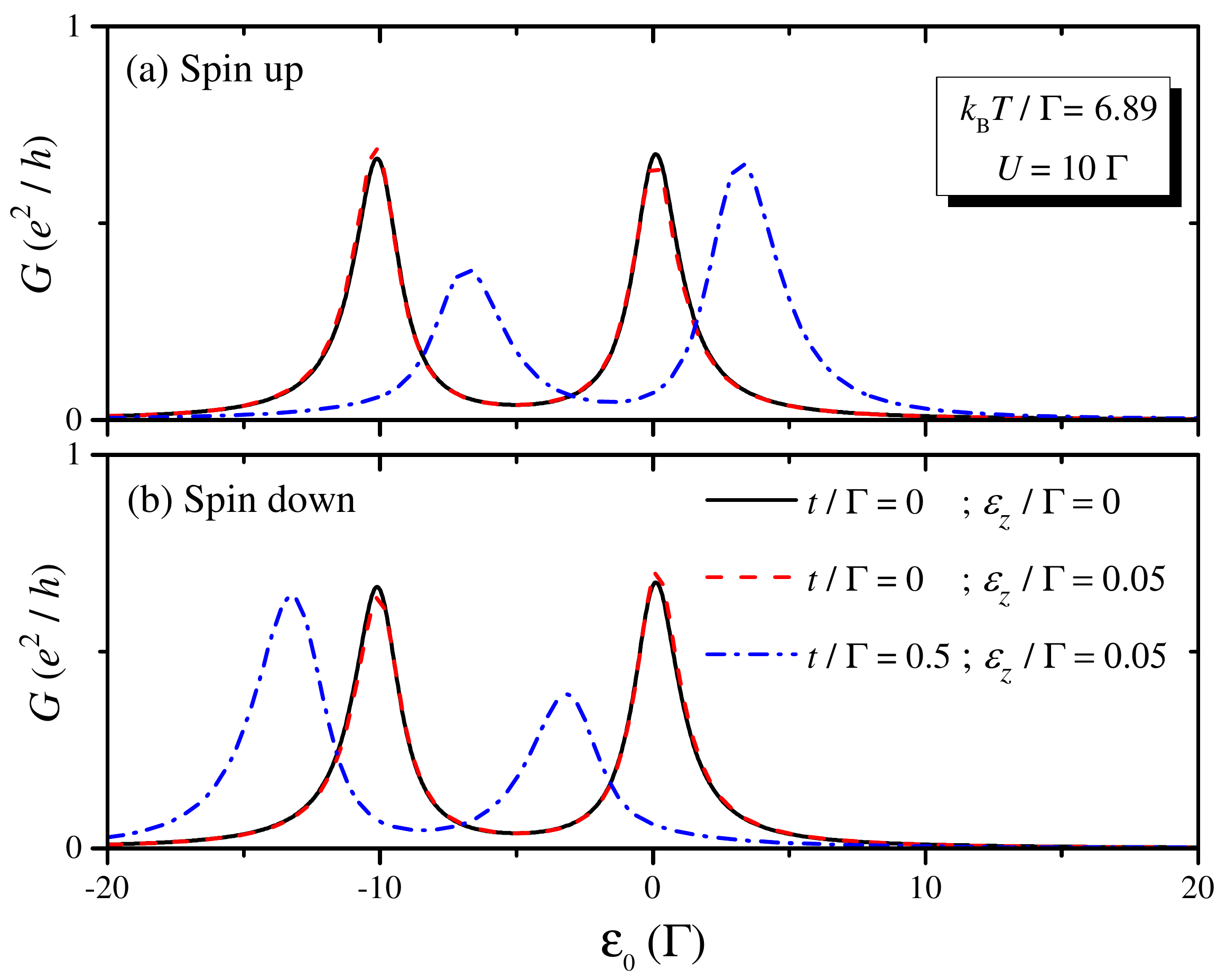}
\caption{Comparison of spin-resolved conductances $G$ at $k_{\text{B}}T/\Gamma=6.89$ [(a) and (b)] for the cases with and without the side-coupled molecule. The single dot case: Without magnetic field (solid black line) and with magnetic field (red dashed line); The case with the side-coupled QDM and magnetic field using $t=\Gamma/2$ (blue dotted-dashed line). In all panels $\mu=0$, $\Delta=\Gamma$ and $U=10\,\Gamma$.}
\label{Fig9}
\end{figure}

\begin{figure}[tbph]
\centering
\includegraphics[width=0.45\textwidth]{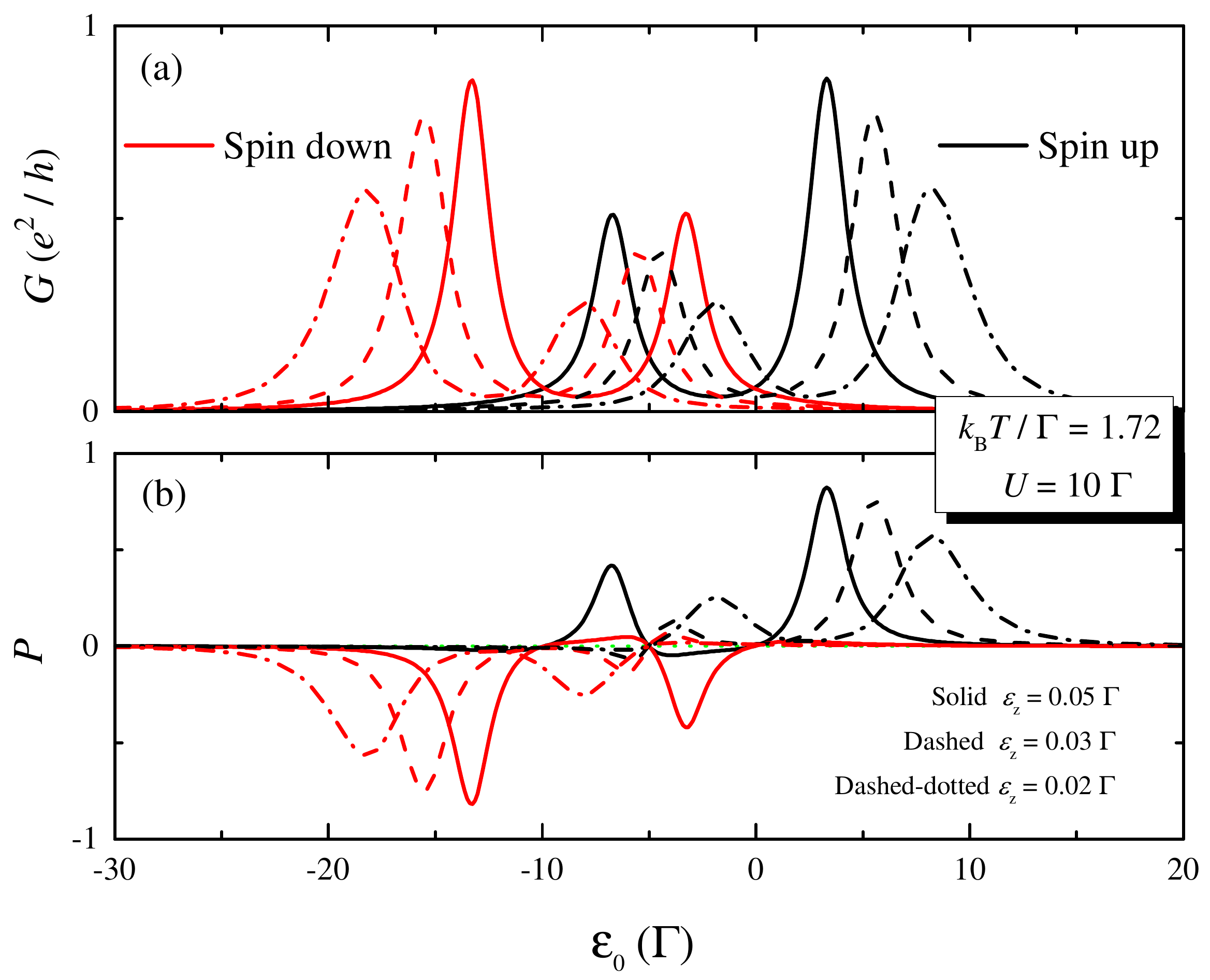}
\caption{Comparison of spin-resolved conductances $G$ (a) and weighted polarization $P$ (b) at $k_{\text{B}}T/\Gamma=1.72$. The magnetic fields are $\ve_{z}=0.05\,\Gamma$ (solid lines), $\ve_{z}=0.03\,\Gamma$ (dashed lines) and $\ve_{z}=0.02\,\Gamma$ (dotted-dashed lines). In all panels black (red) curves are for spin up (down), $\mu=0$, $t=\Gamma/2$, $\Delta=\Gamma$ and $U=10\,\Gamma$.}

\label{Fig10}
\end{figure}

Figure\ \ref{Fig9} shows the conductance as a function of $\varepsilon_0$. In the case with $t=0$ and zero magnetic field (solid black line), we find the standard structure of the Coulomb blockade, \emph{i. e.} two peaks around the energies $\ve_{0}=0$ and $\ve_{0}=-U$. When the magnetic field is turned on with $t=0$ (red dashed line), the position of the peaks of the conductance are shifted by $\ve_{z}$, however, for this value of the magnetic field, this shift is negligible. For the case with $t=\Gamma/2$  (blue dotted-dashed line), the conductance shows two asymmetric peaks. Note that due to the presence of the side attached QDM and the magnetic field, the peaks in the conductance show an important shift. In fact, now the peaks are located around the energies $\ve_{0}\approx (-\ve_{z}-U-\Sigma_{\text{QDM},\uparrow (\downarrow)})$ and $\ve_{0}\approx (-\ve_{z}-\Sigma_{\text{QDM},\uparrow (\downarrow)})$. Clearly, the contribution of the self-energy $\Sigma_{\text{QDM},\uparrow(\downarrow)}$ is negative (positive) because of these two asymmetric peaks move to the right [panel (a)] (left [panel (b)]) respect to the disconnected case, but satisfying the condition $\mid\Sigma_{\text{QDM},\uparrow (\downarrow) }\mid<(\ve_{z}+U)$. This agrees with the noninteracting case (but for $t\neq0$) Fig.\ \ref{Fig4}, and moreover, the value of the highest peak in Fig.\ \ref{Fig9} is equal to the value of a single peak in the Fig.\ \ref{Fig4} for the same parameters under study.

In Fig.\ \ref{Fig10} we further analyze the dependence of spin-resolved conductance of the magnetic field. In Fig.\ \ref{Fig10}(a) we observe that the conductance peaks shifting are strictly sensitive under subtle changes in magnetic field amplitude as a consequence of the connection with the QDM self-energy $\Sigma_{\text{QDM},\sigma}$. Note that the peak shifting is larger as the Zeeman splitting decrease, obtaining zones with nonvanishing conductance for only one spin with a lower magnetic field. By taking advantage of this behavior, we plot in Fig.\ \ref{Fig10}(b) the dimensionless weighted polarization. We note the possibility of a weak magnetic field spin polarizer by tuning $\ve_0$ above or sufficiently below Fermi energy in our system. Note that for $\ve_{0}\sim-U/2$ the conductance is completely non-polarized regardless the magnetic field values. Additionally, we can appreciate that the peaks of the spin-dependent conductance, and consequently, of the spin-dependent polarization, decrease with the magnetic field. This behavior is due to as the magnetic field is reduced, higher values of $\ve_0$ (in absolute value) are needed to carry the system in resonance with the Fermi energy. As $\varepsilon_0$ is increased, the transmission function becomes asymmetrical, and its area is reduced around the Fermi energy, and consequently the peak of the spin-dependent conductance decreases. The previous also holds for the non-interacting case.

\begin{figure}[tbph]
\centering
\includegraphics[width=0.475\textwidth]{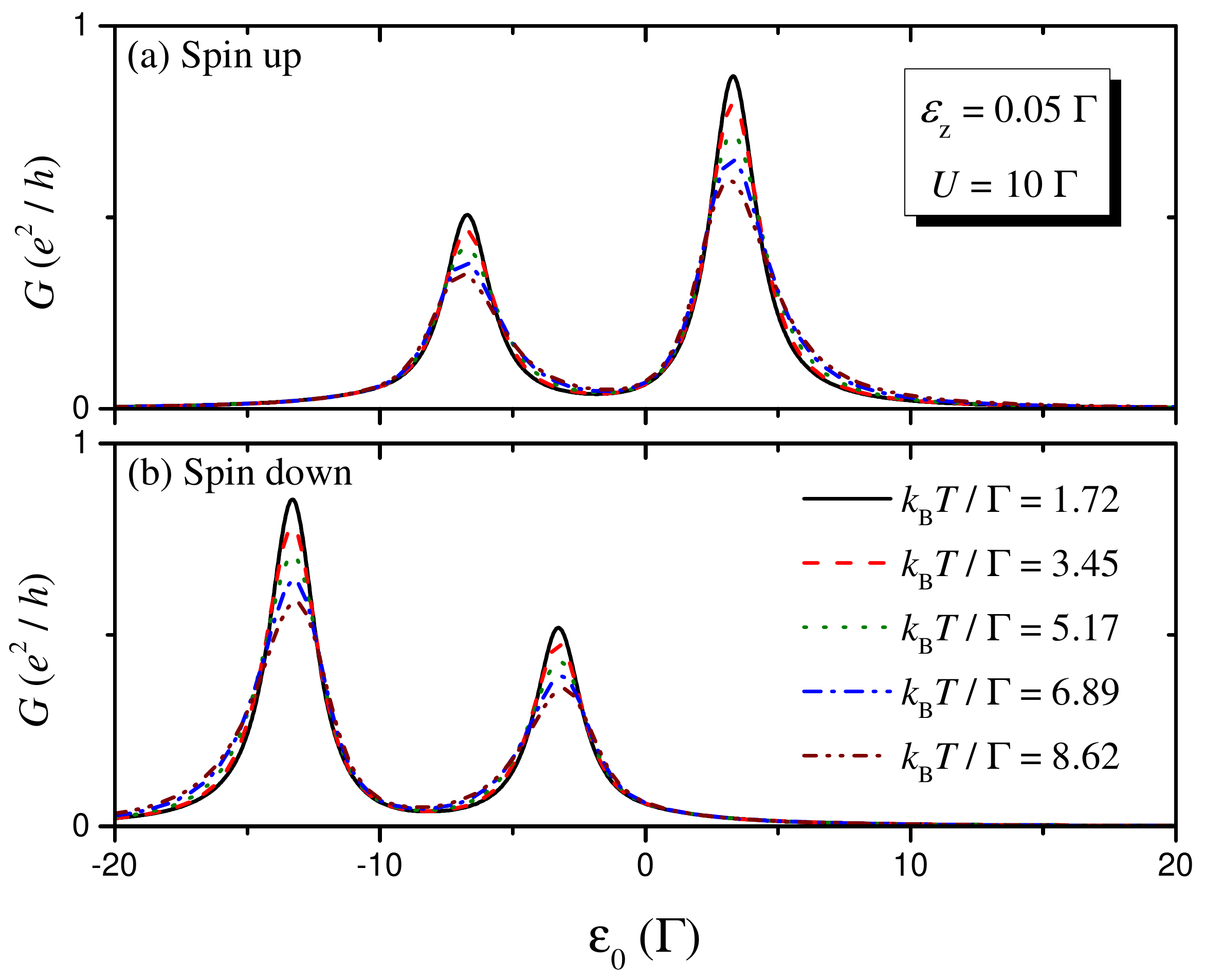}
\caption{Spin-resolved conductance $G$ for different temperatures between 1.72 $\leq k_{\text{B}}T/\Gamma\leq$ 8.62 as a function of $\ve_{0}$. Here the parameters are: $\Delta=\Gamma$, $\ve_{z}=0.05\,\Gamma$, $t=\Gamma/2$ and $\mu=0$.}
\label{Fig11}
\end{figure}

Another important aspect is focused on the conductance dependence on the temperature for the case with $t\neq0$. When we set $t=0$, both in non-interacting and interacting cases, is straightforward to obtain a moderate dependence  of the conductance on the temperature. On the contrary, for $t\neq0$, in Fig.\ \ref{Fig11} the conductance exhibits a notorious dependence of the temperature, by showing a decrease in amplitude at higher temperatures.

\begin{figure}[tbph]
\centering
\includegraphics[width=0.475\textwidth]{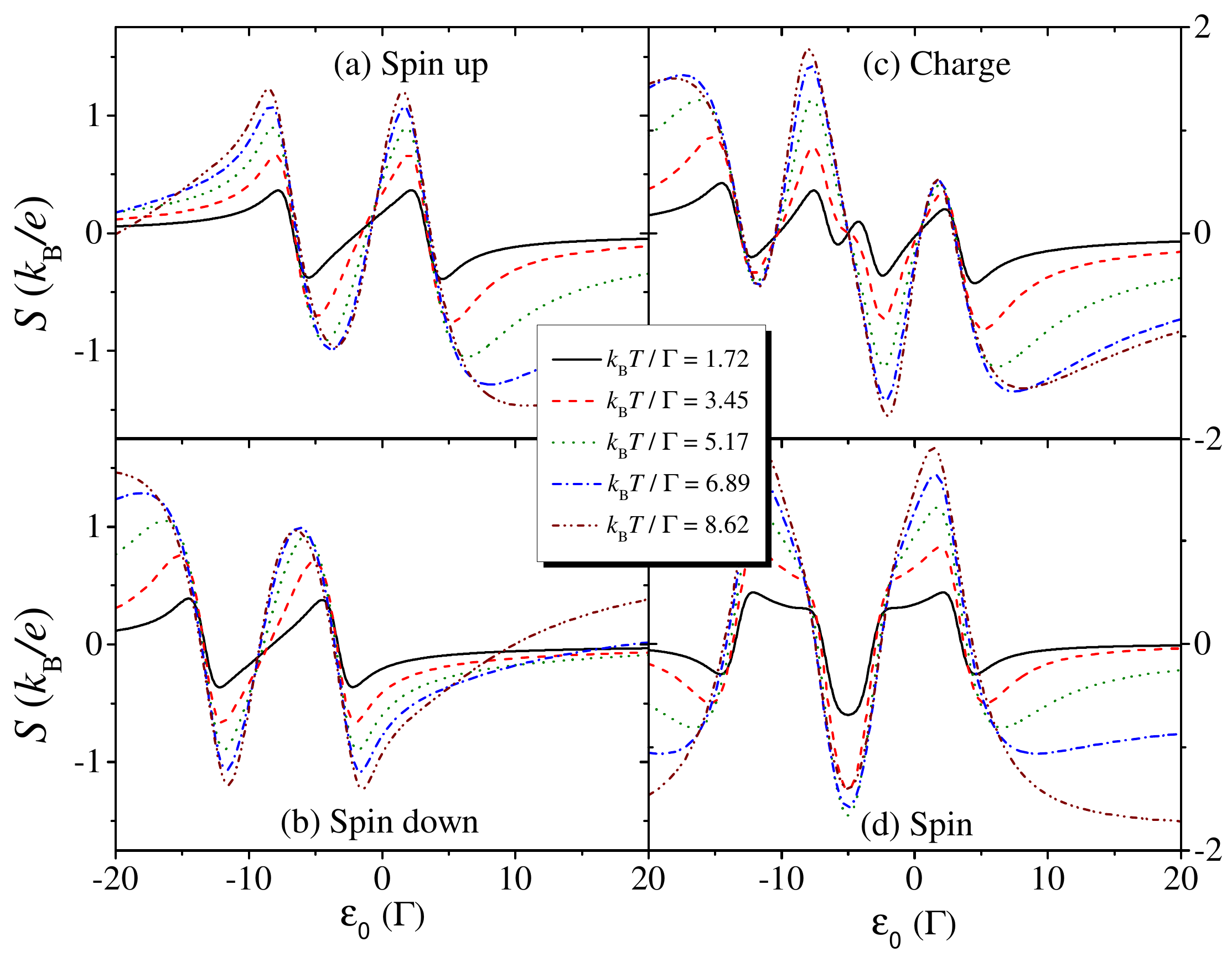}
\caption{Seebeck coefficient $S$ as a function of $\ve_{0}$: Spin-resolved: For spin up (a) and spin down (b); Charge-Seebeck coefficient (c) and spin-Seebeck coefficient (d). Here the parameters are $\Delta=\Gamma$, $\mu=0$ and $t=\Gamma/2$ for different temperatures in the same range used in Fig.\ \ref{Fig11}.}
\label{Fig12}
\end{figure}

\begin{figure}[tbph]
\centering
\includegraphics[width=0.475\textwidth]{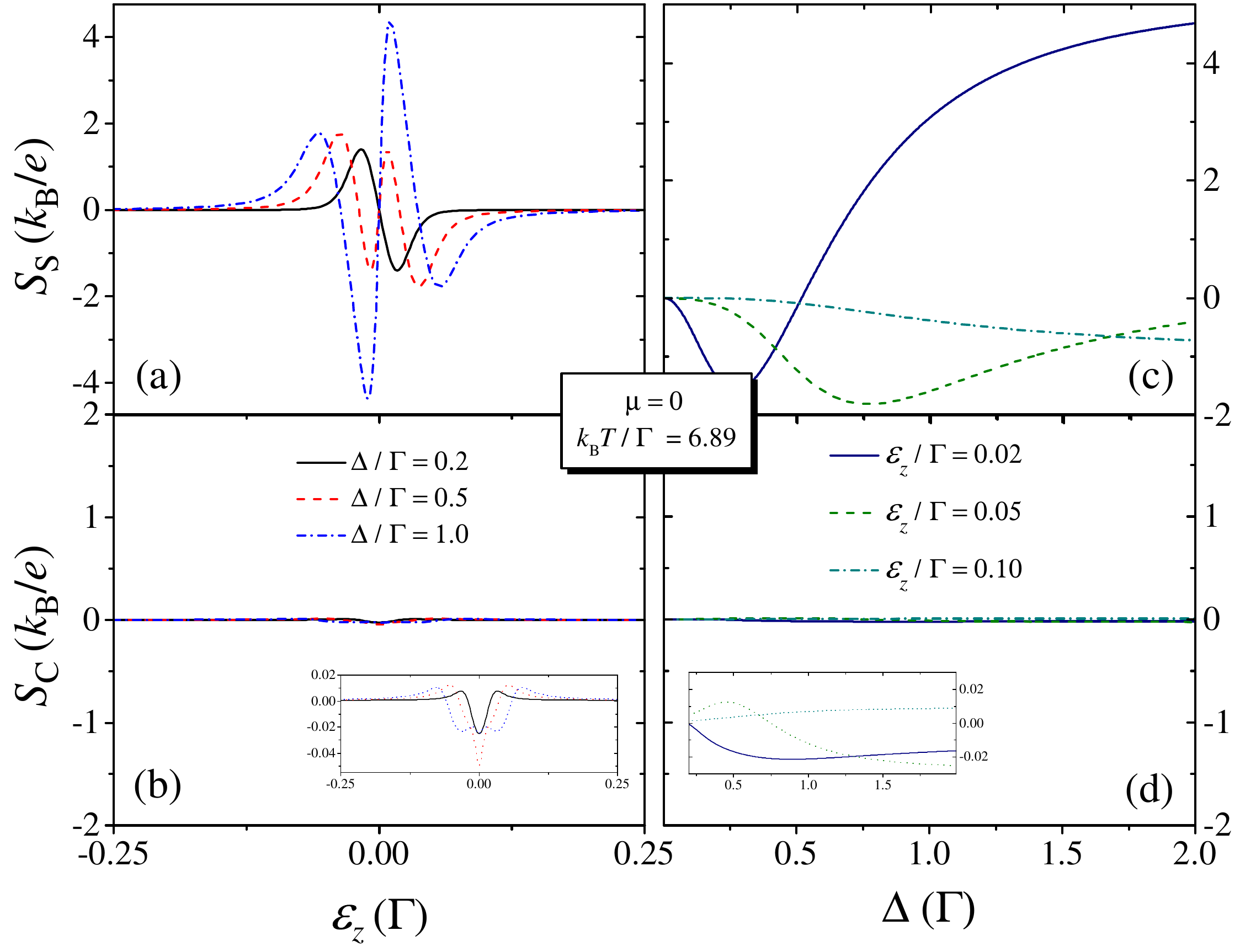}
\caption{Spin-Seebeck $S_{\text{s}}$ and charge-Seebeck $S_{\text{c}}$ coefficients as a function of $\ve_{z}$ [(a) and (b), respectively] and $\Delta$ [(c) and (d), respectively]. Here $k_{\text{B}}T/\Gamma=6.89$, $t=\Gamma/2$, $\mu=0$ and $\ve_0=-4.98\,\Gamma$.}
\label{Fig13}
\end{figure}

Figure\ \ref{Fig12} displays the Seebeck coefficients, per spin, charge, and spin as a function of $\ve_{0}$ for different temperatures. The Seebeck coefficient for both spins vanishes at the resonance points discussed previously. The dependence of the temperature shown in Fig.\ \ref{Fig12} agrees with the one showed before for the conductance. The Seebeck coefficient amplitude must be larger for higher temperatures, contrary to the case of the conductance. In  Fig.\ \ref{Fig12} panels (c) and (d), we also can obtain a pure SSE effect for fixed $\ve_0$, similar to the one described for the non-interacting case. Figure\ \ref{Fig13} shows the charge and spin Seebeck effect as a function of the Zeeman energy. For fixed  $\ve_{0}=-4.98\,\Gamma\sim -U/2$, a pure SSE is reached for a broad range of the Zeeman energy, since $S_{\text{S}}\gg S_{\text{C}}$, being enhanced as $\Delta$ increased [panels (a) and (b)]. On the other hand, Fig.\ \ref{Fig13}(c) displays that the tuning of $S_{\text{S}}$ as function of $\Delta$ can achieve a pure SSE, even for weak magnetic field and since $S_{\text{C}}$ approaches to vanishing [Fig.\ \ref{Fig13}(d)]. Then, it is clear that SSE amplitude increases as more energy asymmetry is considered in the external QDs, as a consequence of the intensified interference phenomena in the system, in analogy to the non-interacting case. It constitutes evidence that even in the presence of interactions in the Coulomb blockade regime a  tunable pure SSE can be reached in our system.  

\section{Summary}\label{secconclu}

We studied the conductance and the Seebeck coefficient between two metallic leads at a temperature difference through a QD with a side-coupled QDM structure. We considered two different cases: without and with Coulomb interaction in the embedded QD. For the first case ($U=0$) without a magnetic field, we controlled the enhancement of the Seebeck coefficient including energy asymmetry in the QDM. The spin-resolved Seebeck coefficient supports a pure SSE by applying a magnetic field with $\mu=0$, $\ve_{0}=0$ and $\Delta\neq0$, for weak magnetic fields. For the second case, in the Coulomb blockade regime, we showed that our system could be used as an efficient spin polarizer device in a wide range of $\ve_0$. Besides, we also obtained a pure SSE by tuning $\ve_{0}$ and holding fixed the other parameters in the system. Moreover, for a fixed $\ve_{0}\sim -U/2$, the effect is achieved in the entire Zeeman energy range considered. In analogy to the charge Seebeck effect which requires breaking particle-hole symmetry, the SSE arises with the combination of particle-hole symmetry breaking and high spin-polarization. In our configuration, a small break of the spin degeneracy produces a high spin-polarization and an enhancement of the spin Seebeck coefficient. Our findings could lead to implementing new exciting spintronic devices, capable to spin-polarized and generating a pure SSE energy conversion for weak magnetic fields.\cite{Hamo2016} 

\begin{acknowledgments}
J.P.R.-A. is grateful for the funding of scholarship CONICYT-Chile No. 21141034. F.J.P. is grateful for the funding of FONDECYT Grant 3170010 and Rosa L\'opez for her instructive discussions. A.G. acknowledges support from DGIP-UTFSM. O.\'A.-O. acknowledges support from NSF Grant No. DMR 1508325. P.A.O. acknowledges support from FONDECYT Grant No. 1140571.
\end{acknowledgments}

\appendix

\section{Embedded quantum dot Green function}

The retarded Green's function in time domain $G^{r}\left(\tau,0\right)$ for fermionic operators $A$ and $B$ is defined as

\begin{equation}
G^{r}_{A,B}\left(\tau\right)= {\la\la A, B \ra\ra}_{\tau}^{r}= -i \theta(\tau) \langle  \left\lbrace A(\tau), B(0)\right\rbrace\rangle,
\end{equation}
whose EOM in energy space takes the following form in the energy domain
\begin{equation}
(\ve+i0^{+}){\la\la A, B \ra\ra}_{\ve}^{r} + {\la\la \left[H,A \right], B \ra\ra}_{\ve}^{r}=\la \left\lbrace A, B \right\rbrace\ra,
\end{equation}
with $H$ the Hamiltonian given in Eq. (\ref{H}) and $0^{+}$ an infinitesimal number. In what follows, we will not write neither the superscript $r$ nor the infinitesimal number $0^{+}$ for simplicity. For QD0, it is straightforward to show that the corresponding Green's function ${\la\la d_{0,\sigma}, d_{0,\sigma}^{\dagger} \ra\ra}_{\ve}$ is given by
\begin{equation}
\begin{aligned}
\left(\ve-\ve_{0,\sigma}\right)& {\la\la d_{0,\sigma}, d_{0, \sigma}^{\dagger}\ra\ra}_{\ve}=1+ U {\la\la d_{0,\sigma} n_{0,\bar{\sigma}},d_{0,\sigma}^{\dagger} \ra\ra}_{\ve} \\ &+ \sum_{k_{\alpha}}\nu_{\alpha}^{*}\la\la c_{k_{\alpha},\sigma}, d_{0 \sigma}^{\dagger}\ra\ra_{\ve} + t\la\la d_{1,\sigma}, d_{0,\sigma}^{\dagger}\ra\ra_{\ve}.
\end{aligned}\label{GR000}
\end{equation}

The equation for ${\la\la c_{k_{\alpha},\sigma}, d_{0,\sigma}^{\dagger}\ra\ra}_{\ve}$ is found to be

\begin{equation}
{\la\la c_{k_{\alpha},\sigma}, d_{0,\sigma}^{\dagger}\ra\ra}_{\ve}=\frac{\nu_{\alpha}}{\left(\ve-\ve_{k_{\alpha}}\right)}\la\la d_{0, \sigma}, d_{0, \sigma}^{\dagger}\ra\ra_{\ve},
\end{equation}
and for $\la\la d_{1,\sigma}, d_{0,\sigma}^{\dagger}\ra\ra$ is
\begin{equation}
\begin{aligned}
\left(\ve-\ve_{1,\sigma}\right)&\la\la d_{1, \sigma}, d_{0, \sigma}^{\dagger}\ra\ra_{\ve}-t^{*}\la\la d_{0,\sigma}, d_{0,\sigma}^{\dagger}\ra\ra_{\ve} \\ = & \ v\left(\la\la d_{2,\sigma}, d_{0,\sigma}^{\dagger}\ra\ra_{\ve} + \la\la d_{3,\sigma}, d_{0,\sigma}^{\dagger}\ra\ra_{\ve}\right).
\end{aligned}\label{GR10}
\end{equation}
In Eq.\ (\ref{GR10}), we use the definition $v_{2}=v_{3}\equiv v$ in order to simplify the discussion. The expressions for the Green's functions $\la\la d_{2,\sigma}, d_{0,\sigma}^{\dagger}\ra\ra$ and $\la\la d_{3,\sigma}, d_{0,\sigma}^{\dagger}\ra\ra$ are given respectively by
\begin{equation}
\la\la d_{2,\sigma}, d_{0,\sigma}^{\dagger}\ra\ra_{\ve}=\frac{v}{\left(\ve-\ve_{2,\sigma}\right)}\la\la d_{1,\sigma},d_{0,\sigma}^{\dagger} \ra\ra_{\ve} \label{GR20}\,,
\end{equation}
\begin{equation}
\la\la d_{3,\sigma}, d_{0,\sigma}^{\dagger}\ra\ra_{\ve}=\frac{v}{\left(\ve-\ve_{3,\sigma}\right)}\la\la d_{1,\sigma},d_{0,\sigma}^{\dagger} \ra\ra_{\ve}. \label{GR30}
\end{equation}

We set $\ve_{2(3),\sigma}\equiv\ve_{1,\sigma}+ (-) \Delta$, so we can combine Eqs.\ (\ref{GR10}), (\ref{GR20}) and (\ref{GR30}) to find
\begin{equation}
\small
\begin{aligned}
\la\la d_{1,\sigma}, d_{0,\sigma}^{\dagger}\ra\ra_{\ve} &=t^{*}\left[\frac{\left(\ve-\ve_{1,\sigma}\right)^{2}-\Delta^{2}}{\left(\ve-\ve_{1,\sigma}\right)\left[\left(\ve-\ve_{1,\sigma}\right)^{2}-\Delta^{2}-2v^{2}\right]}\right] \\ & \times \la\la d_{0,\sigma},d_{0,\sigma}^{\dagger}\ra\ra_{\ve}.
\end{aligned}\label{GR123}
\end{equation}
Replacing the Eq.\ (\ref{GR123}) into Eq.\ (\ref{GR000}) we obtain the form
\begin{eqnarray}
\nonumber
\left(\ve-\ve_{0,\sigma}-\Sigma_{\text{QDM},\sigma}\left(\ve\right)-\Sigma\left(\ve\right)\right)\la\la d_{0,\sigma}, d_{0,\sigma}^{\dagger}\ra\ra_{\ve} \\= 1+ U {\la\la d_{0,\sigma} n_{0,\bar{\sigma}},d_{0,\sigma}^{\dagger} \ra\ra}_{\ve},
\label{A9}
\end{eqnarray}
where we define
\begin{equation}
\Sigma_{\text{QDM},\sigma}\left(\ve\right)=\frac{t^{2}\left[\left(\ve-\ve_{1,\sigma}\right)^{2}-\Delta^{2}\right]}{\left(\ve-\ve_{1,\sigma}\right)\left[\left(\ve-\ve_{1,\sigma}\right)^{2}-\Delta^{2}-2v^{2}\right]},
\end{equation}
and $\Sigma(\ve)$ as
\begin{equation}
\Sigma\left(\ve\right)=\sum_{k_{\alpha}}\frac{|\nu_{\alpha}|^{2}}{\ve-\ve_{k_{\alpha}}}.
\end{equation}
Now we calculate the EOM for the Green function $\la\la d_{0,\sigma}n_{0,\bar{\sigma}},d_{0,\sigma}^{\dagger}\ra\ra_{\ve}$. The complete equation for this case is given by
\begin{equation}
\small
\begin{aligned}
&\left(\ve-\ve_{0,\sigma}-U\right)\la\la d_{0,\sigma}n_{0,\bar{\sigma}},d_{0,\sigma}^{\dagger}\ra\ra_{\ve}+\sum_{k_{\alpha}}\nu_{\alpha}\la\la c_{k_{\alpha},\sigma}^{\dagger}d_{0,\bar{\sigma}}d_{0,\sigma}, d_{0,\sigma}^{\dagger}\ra\ra_{\ve} \\ &- \sum_{k_{\alpha}}\nu_{\alpha
}\la\la d_{0,\bar{\sigma}}^{\dagger}c_{k,\bar{\sigma}}d_{0\sigma}, d_{0, \sigma}^{\dagger}\ra\ra_{\ve}- \sum_{k_{\alpha}}\nu_{\alpha
}\la\la c_{k,\sigma}n_{0,\bar{\sigma}}, d_{0, \sigma}^{\dagger}\ra\ra_{\ve} \\ &+ t\la\la d_{1,\bar{\sigma}}d_{0,\bar{\sigma}}d_{0,\sigma},d_{0,\sigma}^{\dagger}\ra\ra_{\ve} -t\la\la d_{0,\bar{\sigma}}d_{1,\bar{\sigma}}d_{0,\sigma},d_{0,\sigma}^{\dagger}\ra\ra_{\ve} \\ & \  \ \ \  \ \ \ \  -t\la\la d_{1,\sigma}n_{0,\bar{\sigma}},d_{0,\sigma}^{\dagger} \ra\ra_{\ve} \ \ \ = \ \ \ \ \langle n_{0,\bar{\sigma}}\rangle.
\end{aligned}\label{EOM0}
\end{equation}
We keep only the correlations $\la\la c_{k,\sigma}n_{0,\bar{\sigma}}, d_{0, \sigma}^{\dagger}\ra\ra_{\ve}$ and $\la\la d_{1,\sigma}n_{0,\bar{\sigma}},d_{0,\sigma}^{\dagger} \ra\ra_{\ve}$ on the Eq.\ (\ref{EOM0}), so the EOM can be approximated as
\begin{eqnarray}
\nonumber
&\left(\ve-\ve_{0,\sigma}-U\right)\la\la d_{0,\sigma}n_{0,\bar{\sigma}},d_{0,\sigma}^{\dagger}\ra\ra_{\ve} \\ \nonumber &-\sum_{k_{\alpha}}\nu_{\alpha} \la\la c_{k,\sigma}n_{0,\bar{\sigma}}, d_{0, \sigma}^{\dagger}\ra\ra_{\ve}-t\la\la d_{1,\sigma}n_{0,\bar{\sigma}},d_{0,\sigma}^{\dagger} \ra\ra_{\ve} \\ &\approx \langle n_{0,\bar{\sigma}}\rangle.
\label{nnn}
\end{eqnarray}
Under the same approximation treated in the previous case (leaving out the higher correlation terms), we obtain for the Green's function $\la\la c_{k,\sigma}n_{0,\bar{\sigma}}, d_{0, \sigma}^{\dagger}\ra\ra_{\ve}$ the approximate form which is given by
\begin{equation}
\la\la c_{k,\sigma}n_{0,\bar{\sigma}}, d_{0, \sigma}^{\dagger}\ra\ra_{\ve}=\frac{\nu_{\alpha}}{\left(\ve-\ve_{k_{\alpha}}\right)}\la\la d_{0,\sigma}n_{0,\bar{\sigma}},d_{0,\sigma}^{\dagger}\ra\ra_{\ve},
\end{equation}
and for $\la\la d_{1,\sigma}n_{0,\bar{\sigma}},d_{0,\sigma}^{\dagger} \ra\ra_{\ve}$ we obtain
\begin{equation}
\begin{aligned}
& \left(\ve-\ve_{1,\sigma}\right)\la\la d_{1,\sigma}n_{0,\bar{\sigma}},d_{0,\sigma}^{\dagger} \ra\ra_{\ve}-t\la\la d_{0,\sigma}n_{0,\bar{\sigma}},d_{0,\sigma}^{\dagger}\ra\ra_{\ve} \\ &-v\la\la d_{2,\sigma}n_{0,\bar{\sigma}},d_{0,\sigma}^{\dagger}\ra\ra^{r}_{\ve}-v \la\la d_{3,\sigma}n_{0,\bar{\sigma}},d_{0,\sigma}^{\dagger}\ra\ra_{\ve}=0.
\end{aligned}\label{corretodas}
\end{equation}

At this point, is straightforward to obtain the two last Green's functions under this approximation, which are given by

\begin{equation}
\la\la d_{2(3),\sigma}n_{0,\bar{\sigma}},d_{0,\sigma}^{\dagger}\ra\ra_{\ve}=\frac{v}{\left(\ve-\ve_{2(3),\sigma}\right)} \la\la d_{1,\sigma} n_{0,\bar{\sigma}},d_{0,\sigma}\ra\ra_{\ve}\,.\label{corre23}
\end{equation}
Hence, by combining the Eq.\ (\ref{corre23}) with Eq.\ (\ref{corretodas}) and then substituting into Eq.\ (\ref{nnn}), we obtain the Green's function $\la\la d_{0,\sigma}n_{0,\bar{\sigma}},d_{0,\sigma}^{\dagger}\ra\ra_{\ve}$, given by
\begin{equation}
\la\la d_{0,\sigma}n_{0,\bar{\sigma}},d_{0,\sigma}^{\dagger}\ra\ra_{\ve}=\frac{\langle n_{0,\bar{\sigma}}\rangle}{\left[ \ve-\ve_{0,\sigma}-\Sigma\left(\ve\right)-\Sigma_{\text{QDM},\sigma}\left(\ve\right)-U\right]}.
\label{A17}
\end{equation}

Finally, replacing 	Eq.(\ref{A17}) in to Eq.(\ref{A9}), we obtain the EOM for our problem, which is given by

\begin{eqnarray}
\la\la d_{0,\sigma},d_{0,\sigma}^{\dag} \ra\ra_{\ve} = \frac{1-\la n_{\bar{\sigma}}\ra}{\ve-\ve_{0,\sigma}-\Sigma(\ve) -\Sigma_{\text{QDM},\sigma}(\ve)} \nonumber \\ \label{A18}
+\frac{\la n_{\bar{\sigma}}\ra}{\ve-\ve_{0,\sigma}-U-\Sigma(\ve)-\Sigma_{\text{QDM},\sigma}(\ve)}\,.
\end{eqnarray}

For the calculation of self-energy, $\Sigma(\ve)$, we calculate the integral,
\begin{equation}
\begin{aligned}
\Sigma(\ve)&=\int d\epsilon_{k_{\alpha}} \rho\left(\ve_{k_{\alpha}}\right)\frac{\mid \nu(\ve_{k_{\alpha}}) \mid ^{2}}{\ve-\ve_{k_{\alpha}}} \\
&=\mathbb{P} \int d\epsilon_{k_{\alpha}} \rho\left(\ve_{k_{\alpha}}\right)\frac{|\nu(\ve_{k_{\alpha}})|^{2}}{\ve-\ve_{k_{\alpha}}} \\
& -i\pi \rho(\ve)\mid \nu(\ve)\mid^{2}.
\end{aligned}
\end{equation}
Using the wide band approximation, we assume that the product of $\rho(\ve_{k_{\alpha}})\mid\nu_{\ve({k_{\alpha}})}\mid^{2}$ is constant in the band limit, $-D<\ve_{k_{\alpha}}<D$, and therefore
\begin{eqnarray}
    2\pi\rho(\ve_{k_{\alpha}})|\nu(\ve_{k_{\alpha}})|^{2}=\left\{
                \begin{array}{ll}
                  \Gamma, \ \mbox{for} -D<\ve_{k_{\alpha}}<D \\
                  0, \ \mbox{for} \ D<|\ve_{k_{\alpha}}|.
                \end{array}
              \right.
 \end{eqnarray}
Under this consideration, we obtain,
\begin{equation}
\begin{aligned}
\Sigma(\varepsilon)\approx - \frac{\Gamma}{\pi}\ln\left|\frac{D+\ve}{D-\ve}\right|-i\frac{\Gamma}{2}.
\end{aligned}
\end{equation}
Therefore, we can reduce the expression for the Eq.(\ref{A18}) in the form
\begin{eqnarray}
\la\la d_{0,\sigma},d_{0,\sigma}^{\dag} \ra\ra_{\ve} = \frac{1-\la n_{\bar{\sigma}}\ra}{\ve-\tilde{\ve}_{0,\sigma}+i\tilde{\Gamma}-\Sigma_{\text{QDM},\sigma}(\ve)} \nonumber \\ \label{A22}
+\frac{\la n_{\bar{\sigma}}\ra}{\ve-\tilde{\ve}_{0,\sigma}-U+i\tilde{\Gamma}-\Sigma_{\text{QDM},\sigma}(\ve)}\,,
\end{eqnarray}
where, we define $\tilde{\Gamma}=\Gamma/2$ and $\tilde{\ve}_{0,\sigma}=\ve_{0,\sigma}+\mathbb{R}e \left(\Sigma(\ve)\right)$. It is important to recall that the function that we obtain in $\mathbb{R}e \left(\Sigma(\ve)\right)$ varies slowly (with respect to energy) and in the integrals can be considered as a constant. Throughout this work, we do not use the notation $\tilde{\Gamma}$ and $\tilde{\ve}_{0,\sigma}$ in order to avoid an overpopulated notation in the manuscript.

Lastly, the spin-dependent transmission probability defined in Eq.\ (\ref{tras}) is explicitly given by
\begin{widetext}
\begin{equation}
\mathcal{T}_{\sigma}(\ve)=\Gamma^2\left[\frac{(\ve-\ve_{0,\sigma}-U-\Sigma_{\text{QDM},\sigma})^2+\Gamma^2-U\langle n_{\bar{\sigma}}\rangle(U-2(\ve-\ve_{0,\sigma}-\Sigma_{\text{QDM},\sigma}))}{((\ve-\ve_{0,\sigma}-U-\Sigma_{\text{QDM},\sigma})^2+\Gamma^2)((\ve-\ve_{0,\sigma}-\Sigma_{\text{QDM},\sigma})^2+\Gamma^2)}\right]\,.
\end{equation}
\end{widetext}
\noindent

\bibliography{biblio}
\bibliographystyle{apsrev4-1}

\end{document}